\documentclass[12pt,a4paper, final]{iopart}
\usepackage{color}
\usepackage{hyperref}
\usepackage{amssymb}
\usepackage{gensymb}
\usepackage{eufrak}
\usepackage{setspace}
\usepackage{subfigure}
\usepackage{import}
\usepackage{float}
\usepackage{graphicx}
\usepackage{overpic}
\usepackage{xspace}
\usepackage[normalem]{ulem}

\usepackage{tikz}
\usetikzlibrary{shapes,arrows}
\usetikzlibrary{decorations.markings}

\newcommand{\be}{\begin{equation}}
\newcommand{\ee}{\end{equation}}

\newcommand{\sinc}{\mathop{\mathrm{sinc}}\nolimits}
\renewcommand{\e}{\mathrm{\rm e}}
\renewcommand{\d}{\,{\mathrm{d}}}
\newcommand{\p}{\,{\textrm{p}}}

\newcommand{\defeq}{\mathrel{\mathop:}=} 
\newcommand{\abs}[1]{\left| {#1} \right|} 

\begin{document}
\title[Spectral Interpolation for detection of continuous gravitational waves]{A targeted spectral
interpolation algorithm for the detection of continuous gravitational waves}
\author{Gareth~S.~Davies$^{1,2}$, Matthew~Pitkin$^1$ \& Graham~Woan$^1$}
\address{$^1$ SUPA, University of Glasgow, Glasgow, G12 8QQ, United Kingdom}
\address{$^2$ Cranfield Forensic Institute, Cranfield University, Shrivenham, SN6 8LA, United Kingdom}
\ead{\mailto{g.s.davies@cranfield.ac.uk}}

\begin{abstract} We present an improved method of targeting continuous gravitational-wave signals in data from
the LIGO and Virgo detectors with a higher efficiency than the time-domain Bayesian pipeline used in many
previous searches. Our spectral interpolation algorithm, SplInter, removes the intrinsic phase evolution of
the signal from source rotation and relative detector motion. We do this in the frequency domain and generate
a time series containing only variations in the signal due to the antenna pattern. Although less flexible
than the classic heterodyne approach, SplInter allows for rapid
analysis of putative signals from isolated (and some binary) pulsars, and efficient follow-up searches for
candidate signals generated by other search methods. The computational saving over the heterodyne approach can
be many orders of magnitude, up to a factor of around fifty thousand in some cases, with a minimal impact on
overall sensitivity for most targets.
\end{abstract}

\maketitle

\section{Introduction}
\label{Sec:Intro}
Rapidly rotating neutron stars are promising sources of long-lived gravitational-wave signals and one of the
key science targets of the LIGO and Virgo Scientific Collaborations \cite{KnownPulsarsS6}. The full parameter
space for these signals is too large for simple coherent methods to be employed on timescales longer than a few
days, so a range of more specific methods have been developed to explore specific regions of the space to
different depths \cite{2010ApJ...713..671A}. Known radio and X-ray pulsars comprise an important class of
potential gravitational-wave source and three analysis pipelines have been specifically developed to exploit
the known rotational phase evolution of these targets \cite{RJDandWoan,5nvector,2010CQGra..27s4015J}. These
targeted pipelines are fully-coherent over arbitrary lengths of time, tracking the predicted gravitational
signal based on electromagnetic observations. They perform the deepest gravitational wave searches in the field
and all three use both time and frequency domain techniques to reduce the data at relatively low computational
cost. However, as these pipelines are now being applied more widely to candidate sources identified by other
searches there is significant benefit in reducing the cost still further. Two of these (the Bayesian targeted
and $\mathcal{G}$ statistic \cite{RJDandWoan,2010CQGra..27s4015J} pipelines), rely on data from a carefully
implemented, but slow, heterodyne step developed by Dupuis and Woan
\cite{2010ApJ...713..671A,RJDandWoan,RJDThesis,2007PhRvD..76d2001A} that allows the data to be sampled at a
much lower rate than is generated by the detector (from $\sim 16$\, kHz to usually one sample per minute).
Although we believe this exact solution is still the best way to approach high-value targets, short-period binaries and
targets close to spectral lines (such as the Crab pulsar), when certain approximations are valid the step can
be performed more efficiently for many
other targets using fast fourier transform (FFT) methods. We therefore present an efficient method for
down-sampling gravitational wave data and removing the effects of detector motion with respect to the source
based on FFTs. Similar spectral methods have been used widely in the field for many years, and indeed form the
basis of the $\mathcal{F}$-statistic search methods \cite{JKBook} which are ubiquitous.  Our version of spectral interpolation, 
\textit{SplInter}, is designed as a more efficient replacement for the heterodyne algorithm in certain situations. As we will show, 
our algorithm's large computational costs savings and very small sensitivity losses (when certain signal assumptions are valid) mean 
that it can be quickly and easily applied to a large number of targets, e.g. if following up large numbers of potential candidate signals from blind all-sky searches.

In \sref{Sec:SplInter} we show how to calculate the down-sampled data streams using this method. In
\sref{Sec:SplTest} we confirm the equivalence of these streams to those from the heterodyne method, and assess
the improvement in computational efficiency offered by SplInter.

The continuous gravitational wave strain signal in the output of a detector depends on the source emission
mechanism and the source/detector geometry, but for the purposes of this analysis we assume it to be
quasi-sinusoidal, with the form
\be
h(T) = A(T)\e^{\rmi\Phi(T)}+A^*(T) \e^{-\rmi\Phi(T)},
\label{eqn:hoft}
\ee
where $A(T)$ contains both the antenna response pattern and source amplitude parameters. $\Phi(T)$ is the
apparent phase evolution of the signal due to source rotation and $T$ is the time in a suitable inertial frame
(see below). For example in the case of a triaxial neutron star rotating about a principle axis, emitting
gravitational waves at twice the rotation frequency, $A$ has the form \cite{RJDandWoan}
\be
A(T) = \frac{1}{4} F_+(T;\psi) h_0 (1+\cos^2\iota) - \frac{\rmi}{2} F_\times(T;\psi)h_0 \cos \iota,
\label{eqn:Aoft}
\ee
where $\iota$ is the inclination angle between the source rotation axis and the line of sight from the
detector to the source, $\psi$ is the gravitational wave polarisation angle and $h_0$ is the wave amplitude.
$F_+$ and $F_\times$ are the antenna pattern responses to plus and cross polarisations  respectively. The
evolution of $\Phi(T)$ depends on the intrinsic rotational evolution of the neutron star, defined by its
frequency and frequency derivatives $f^{(0),(1),(2)...}_\textrm{\footnotesize rotation}$. Over short
timescales, the time-dependence of $h(T)$ is dominated by this phase term, expanded as
\be
\Phi(T) = \Phi_0 + 2\pi\sum_{l=0} \frac{f^{(l)} (T-T_0)^{l+1}}{(l+1)!},
\label{eqn:phaseT}
\ee
where $f^{(l)}$ is the  $l^{\rm th}$ time derivative of the \emph{gravitational wave} frequency (note that
these are twice the rotation values for the  $l=m=2$ harmonic of a non-precessing, triaxial neutron star),
$T_0$ is the epoch at which ${\Phi(T_0)=\Phi_0}$. The $f^{(l)}$ values are derived from radio,  X-ray, or
$\gamma$-ray pulse times-of-arrival, preferably from data spanning the same analysis period as the
gravitational wave search in question.

The rotational and orbital motions of the Earth put the detector in a non-inertial rest frame, but for a given
source position on the sky we can relate the topocentric signal arrival times at the detector, $t'$, to
those in the source's frame of reference by
\be
T = t' + \delta (t'),
\label{eqn:timeconvert}
\ee
where $\delta (t')$ comprises four time-delay terms \cite{RJDandWoan}:
\be
\delta (t')=\Delta_{\textrm{R}\odot} + \Delta_{\textrm{S}\odot} + \Delta_{\textrm{E}\odot} + \Delta_{\textrm{\footnotesize Binary}}.
\label{eqn:deltat}
\ee
The Roemer delay $\Delta_{\textrm{R}\odot}$ is the dominant term; it is the Euclidian difference in time taken
for the signal to arrive at the detector and the solar system barycentre (SSB). $\Delta_{\textrm{S}\odot}$ is
the Shapiro delay, caused by the bending of spacetime near to massive bodies, which in the case of an
Earth-based detector is dominated by the Sun's contribution. The Einstein delay $\Delta_{\textrm{E}\odot}$
combines the effects of special and general relativistic time dilation due to motion and the presence of
massive bodies. All of these terms vary slowly over the course of a year, and by small amounts over the course
of a day, and as such can be considered as changing only linearly over the half-hour intervals we will
consider shortly. These effects can be addressed in a number of ways, including resampling \cite{PhysRevD.81.084032} 
or heterodyning \cite{RJDandWoan} the data. $\Delta_{\rm Binary}$ however can vary more quickly, on the timescale of the period of
binary motion, this is an additional all-encompassing term that combines the Roemer, Shapiro and Einstein
delays caused by the source's non-inertial motion, should it be in a binary system. In terms of the
topocentric time we now have
\be
\Phi(t') = \phi_0 + 2\pi\sum_{l=0} \frac{f^{(l)} (t'+\delta(t) - t_0)^{l+1}}{(l+1)!},
\label{eqn:phase}
\ee
where $t_0$ is the time at which $\Phi(t_0-\delta) = \phi_0$.

Any difference in the assumed and actual phase evolution would introduce a residual phase evolution 
and a reduction in final search sensitivity. Typically, known radio and X-ray pulsars are timed sufficiently well 
for these effects to be negligible, but they can be included straightforwardly in the subsequent parameter 
estimation stages of a search. Once we have corrected for this known phase variation the only remaining time-dependence in
Equation~\ref{eqn:hoft} is from $A(t)$, which evolves as the antenna pattern sweeps over the source in a
sidereal day. We can therefore sample the data at a much reduced rate, limited only by the changing antenna
pattern, provided that we still capture this. The original heterodyne pipeline achieves phase correction by
multiplying the strain time series $s(t')$ (where we use the notation of \cite{RJDandWoan} that $s(t') =
h(t') + n(t')$, where $h(t')$ is the signal from Equation~\ref{eqn:hoft} and $n(t')$ is Gaussian noise) by
$\exp[ - \rmi\phi(t')]$, where $\phi(t')=\Phi(t')-\phi_0$,
effectively shifting the signal frequency to zero and leaving $A(t)$ as the only time-dependent term.  After
applying a low-pass filter (conventionally with a time constant of 1 minute) we average over $M$ data samples
to obtain a down-sampled time series of the form
\be
B_K = \frac{1}{M}\sum\limits_{j=1}^M s(t'_j) \exp[-\rmi\phi(t'_j)],
\ee
where $M$ is the number of raw data samples contributing to $B_K$ (following \cite{RJDandWoan} this is often
chosen to give one sample per minute), and $K$ is the time index for the resulting
time series. We model this as a combination of a signal term $y_K = A(t_K)\exp(\rmi\phi_0) $ and,
appropriately over the narrow bands considered here,  white Gaussian noise $n_K \sim N(0,\sigma_K^2)$
\be
B_K = y_K+n_K.
\label{eqn:Bkeqykplnk}
\ee
$\sigma_K$ is modelled as constant over short timescales, and is related to the original time-series noise $\sigma_T$ by
\be
\sigma_K^2 = \frac{\sigma_T^2}{M} = \frac{\sigma_T^2}{r\Delta t},
\label{eqn:noiseTDtokconvert}
\ee
where $r$ is the original sample rate and $\Delta t$ is the down-sampled period.

\section{Formulation in the frequency domain}
\label{SubSec:FourierTransform}
We can consider a similar analysis in the frequency domain. The Fourier transform of a signal $h(t')$ limited
in duration to $\Delta t$ centred on a time ($t_k-t_0$) is
\begin{equation}
\fl H_k(f) = \int\limits^{t_k-t_0+\Delta t / 2}_{t_k-t_0-\Delta t / 2}
\left[A(t')\e^{\rmi\Phi(t')}+A^*(t')
       \e^{-\rmi\Phi(t')}\right]\exp\left[-2\pi \rmi f(t'-t_0-t_k+\Delta t / 2)\right] \d t'.
\nonumber\\
\label{eqn:HoffWrecPhi}
\end{equation}
In this Fourier-based version,  $t_k$ become the time-stamps of a series similar to the $t_K$ series defined
above, with $H_k(f_{\rm signal})$ playing the role of $y_K$. However, we now use $k$ rather than $K$ as the
index to highlight that the two sampling rates need not be (and indeed usually are not) the same. $t_0$ is
(again) the reference epoch of our timing solution.

We now consider $\phi(t')$, the time-dependent part of $\Phi(t')$, and use a time coordinate $t$ with its
origin at the mid-point of the data under consideration i.e., ${t' \rightarrow t = t' - t_k+t_0}$.
Importantly, the time-delay correction term $\delta (t)$ in \eref{eqn:deltat} will vary slightly over the time
$\Delta t$. We approximate these changes in the arrival time to first order in $t$, defining $\delta_k =
\delta (t_k)$ and $\dot{\delta}_k = \frac{\d}{\d t}\delta (t_k)$, such that for the duration of the data
${\delta (t) \approx \delta_k + \dot{\delta}_kt}$. Equation \ref{eqn:phase} now becomes
\be
\phi(t) \approx 2\pi  \sum_{l=0} \frac{f^{(l)} \left[t(1+\dot{\delta}_k) + t_k - t_0 +\delta_k\right]^{l +1}}{(l+1)!}.\nonumber
\ee
To second order in $t$
\be
\phi(t)\approx \phi_k + 2\pi f_k t + \pi \dot{f_k} t^2,
\label{eqn:phaseApprox}
\ee
where
\be
\phi_k \defeq 2\pi \sum_{l=0} \frac{f^{(l)}(t_k -t_0 + \delta_k)^{l+1}}{(l +1)!},
\label{eqn:phiK}
\ee
and
\be
f_k \defeq (1+\dot{\delta}_k) \sum_{l=0} \frac{f^{(l)}(t_k -t_0 + \delta_k)^l }{l!}.
\label{eqn:fk}
\ee
We also approximate $\dot{f}_k$ as
\be
\dot{f}_k = \frac{f_{\rm end} - f_{\rm start}}{\Delta t}.
\label{eqn:fdotk}
\ee
where
\be
f_{\rm start} = \left[1+\frac{\d}{\d t}\delta\left(t_k-\frac{\Delta t}{2}\right) \right] \sum_{l=0} \frac{f^{(l)}\left[t_k - \frac{\Delta t}{2}-t_0 + \delta\left(t_k-\frac{\Delta t}{2}\right)\right]^l }{l!}.
\label{eqn:fstart}
\ee
and
\be
f_{\rm end} = \left[1+\frac{\d}{\d t}\delta\left(t_k+\frac{\Delta t}{2}\right) \right] \sum_{l=0} \frac{f^{(l)}\left[t_k + \frac{\Delta t}{2}-t_0 + \delta\left(t_k+\frac{\Delta t}{2}\right)\right]^l }{l!} .
\label{eqn:fend}
\ee
We assume the signal amplitude and antenna pattern contributions to $A(t)$ are approximately constant on
timescales of a small fraction of a day, so when $\Delta t$ is small we can replace $A(t)$ above with
$A(t_k)$. Having defined ${y_k:=A(t_k)\e^{\rmi\phi_0} }$ we can therefore write, using these approximations,
\be
H_k(f) \approx \e^{-\rmi\pi f\Delta t}\int\limits^{\Delta t / 2}_{-\Delta t / 2} \left[y_k \e^{\rmi\phi(t)}+y_k^* \e^{-\rmi\phi(t)}\right]
\e^{-2\pi\rmi ft} \d t. \label{eqn:Hoffyk}
\ee
This is the Fourier transform which will be considered in the following models.

\subsection{The form of the signal in a short transform}
\label{SubSec:SplAlgModels}
Our signal is quasi-sinusoidal, but with amplitude and phase varying slowly as the source moves though the
antenna pattern, and with changes in delay and doppler shift as well as intrinsic variations in the source
spin rate. To first order in $f$, using $H_k(f)$ from \eref{eqn:Hoffyk} and $\phi(t)$ from
\eref{eqn:phaseApprox} we have
\begin{eqnarray}
H_k(f) \approx y_k&\e^{\rmi\phi_k-\rmi\pi f\Delta t }\int\limits^{\Delta t / 2}_{-\Delta t / 2}
\exp \left[2\pi\rmi (f_k-f) t + \rmi\pi \dot{f_k} t^2\right] \d t \nonumber \\
&+ y_k^*\e^{-\rmi\phi_k-\rmi\pi f\Delta t }\int\limits^{\Delta t / 2}_{-\Delta t / 2} \exp
\left[-2\pi\rmi (f_k+f) t - \rmi\pi \dot{f_k} t^2\right] \d t.
\label{eqn:HoffykSplit}
\end{eqnarray}
These expressions are not strictly analytic due to the $t^2$ phase-dependency of the exponent, but are forms
of the familiar Fresnel integral. The limiting form, when $\dot{f_k}$ is small, is just the Fourier transform
of a time-limited sinusoid, so we will consider this as a special case below.

\subsubsection{The sinc approximation, $\dot{f_k} = 0$}
The contribution of the intrinsic spin-down of the source, $f^{(1)}$, to the overall $\dot{f}$ is generally
negligible over the course of a single transform (maybe lasting minutes to hours) and the change in frequency
due to this component will be much smaller than the frequency resolution. For example the Crab pulsar, which
has an unusually large spin-down of $f^{(1)}=7.4\times10^{-10}$\,Hz\,s$^{-1}$ \cite{ATNF, Manchester:2005},
will change in frequency by only $\Delta f = 1.3\times 10^{-6}$\,Hz over 30 minutes, which is $0.1\%$ of the
width of a frequency bin. Instead, $\dot{f}_k$ is dominated by the $\dot{\delta}_k$ term caused by the orbital
motions of the source and observer. The spin and orbital motion of the observer are also usually negligible
over $\sim 1$\,h, so for a source that is \emph{not} in a binary system we can assume $\dot{f}_k = 0$ for the
duration of the integral, so that
\begin{eqnarray}
H_k(f) &\approx y_k\e^{\rmi\phi_k -\rmi\pi f\Delta t }\int\limits^{\Delta t / 2}_{-\Delta t / 2}
\exp \left[2\pi\rmi (f_k -f)t\right] \d t \nonumber\\
&\qquad + y_k^*\e^{-\rmi\phi_k-\rmi\pi f\Delta t}\int\limits^{\Delta t / 2}_{-\Delta t / 2} \exp
\left[- 2\pi\rmi (f_k + f)t\right] \d t\\
\label{eqn:HoffSincPreCalc}
&\approx y_k\Delta t \exp\left[\rmi\phi_k -\rmi\pi f\Delta t\right] \sinc \left[ \pi (f_k-f)\Delta
t\right] \nonumber\\
&\qquad + y_k^*\Delta t \exp\left[-\rmi\phi_k -\rmi\pi f\Delta t\right] \sinc \left[\pi
(f_k+f)\Delta t\right],
\label{eqn:HoffSincincnegfreq}
\end{eqnarray}
where we use the convention
\be
\sinc x \defeq \frac{\sin x}{x}.
\label{eqn:sincDefine}
\ee
Close to the signal frequency (i.e., when $f \simeq f_k$)
\be
\sinc \left[ \pi (f_k-f)\Delta t\right] \gg \sinc \left[ \pi (f_k+f)\Delta t\right],
\ee
so we can remove the second term in \eref{eqn:HoffSincincnegfreq} to give
\be
H_k(f) \approx y_k\Delta t  \exp\left[\rmi\phi_k -\rmi\pi f \Delta t\right] \sinc \left[ \pi
(f_k-f)\Delta t\right].
\label{eqn:HoffSinc}
\ee
We will refer to this below as the \textit{sinc approximation}.
\subsubsection{The Fresnel approximation}
If $\dot{f}_k$ is non-negligible then we can still approximate the form of $H_k(f)$ through a numerical
integration. Such circumstances would occur if the doppler-shifted frequency was evolving significantly on
timescales of $\Delta t$ due to the orbital motion of the source or observer. If the rate of change of signal
frequency is a constant, i.e. $\ddot{f}_k = 0 $, we would expect the signal to appear as a `Fresnel' pattern
in the Fourier transform, characterised by the quadratic evolution of phase with time. Fresnel Integrals have been
studied extensively, and there are good algorithms for fairly rapid calculation \cite{NumRecWeb}. They
comprise a pair of functions defined as \cite{GrahamsBook}
\be
\mathcal{C}[w] := \int\limits_0^w \cos\left(\frac{\pi x^2}{2}\right) \d x
\label{eqn:FresnelC}
\ee
and
\be
\mathcal{S}[w] := \int\limits_0^w \sin\left(\frac{\pi x^2}{2}\right) \d x.
\label{eqn:FresnelS}
\ee
In terms of these integral functions, \eref{eqn:HoffykSplit} becomes
\be
\fl H_k(f) \approx  \frac{y_k}{2\sqrt{2|\dot{f}_k|}}\exp(\rmi\Delta \phi)\bigg\{\mathcal{C}[w_{\rm
end}] +
\rmi\frac{\dot{f}_k}{|\dot{f}_k|}\mathcal{S}[w_{\rm end}] - \mathcal{C}[w_{\rm start}] -
\rmi\frac{\dot{f}_k}{|\dot{f}_k|}\mathcal{S}[w_{\rm start}]\bigg\},
\label{eqn:HoffykFresnel}
\ee
where
\be
\Delta \phi = \phi_k - \pi f\Delta t + \pi \frac{\dot{f}_k \Delta t^2}{2} - \frac{\pi}{\dot{f}_k}\left(f_k + \frac{\dot{f}_k\Delta t}{2} - f \right)^2,
\ee
\be
w_{\rm end} = \frac{\sqrt{2|\dot{f}_k|}}{\dot{f}_k}\left( f_k - f + \frac{\dot{f}_k \Delta t}{2}\right) + \sqrt{\frac{|\dot{f}_k|}{2}}\Delta t
\ee
and
\be
w_{\rm start} = \frac{\sqrt{2|\dot{f}_k|}}{\dot{f}_k}\left( f_k - f + \frac{\dot{f}_k \Delta t}{2}\right) - \sqrt{\frac{|\dot{f}_k|}{2}}\Delta t.
\ee
Here, we have again ignored the $(f+f_k)$ term in \eref{eqn:HoffykSplit}, as again it is negligible in the
interpolation region where $f \simeq f_k$.

We will refer to \eref{eqn:HoffykFresnel} as the \textit{Fresnel approximation} to the signal Fourier transform, and we
calculate the Fresnel integral terms with sufficient numerical precision using the algorithm in
\cite{NumRecWeb}. For small $\dot{f}_k$ the Fresnel approximation reduces to the sinc approximation described
above. Computationally it is more expensive than the sinc approximation, however it need only be implemented
during periods of time corresponding to large values of $\dot{f}_k$,, i.e.,  $|\dot{f}_k|\Delta t^2 > 0.1$.

\section{The Spectral Interpolation algorithm}
\label{Sec:SplInter}
The Spectral Interpolation algorithm (SplInter) is an alternative to the time-domain heterodyne algorithm of
Dupuis and Woan~\cite{RJDandWoan} originally developed for gravitational-wave searches targeting known
pulsars. In contrast to this heterodyne algorithm, SplInter uses Fast Fourier Transform methods to generate a
similarly narrow-band time series but can process multiple targets very much more efficiently and usually with
an acceptable impact on overall sensitivity.  Within the LIGO Scientific Collaboration a Fourier transform data
format, known as `Short-time Fourier Transforms' (SFTs\label{abbrev:SFTs}), has been defined
\cite{SFTs, 2004PhRvD..69h2004A} for use in a variety of continuous gravitational wave searches. These SFTs
contain discrete Fourier transforms of windowed data segments that are much shorter than the duration of the
experiment (usually around half an hour). Of course there is an
associated computational load in creating these SFTs, but this is offset by the efficiency of the SplInter
algorithm. In addition, SFTs for several types of continuous-wave search (such as \cite{GeneralisedPowerFlux})
can be shared with SplInter.

In the first stage of the SplInter algorithm we take a discretely-sampled Fourier transform, in the form of an
SFT, and compute a value of $H_k$ at the instantaneous topocentric signal frequency using one of the
interpolation methods described above over a small number of spectral points either side of the central
topocentric frequency bin. We denote the result of this spectral interpolation $B_k$. In addition, we wish to
calculate $\sigma_k$, the standard deviation of the noise on our estimate of $B_k$.

\subsection{$B_k$ and $\sigma_k$ calculation} \label{SubSec:SplAlgBkEst}
An SFT is of course a discrete Fourier transform, so we must interpolate between frequency bins to recover an
unbiased estimate, $B_k$, of the signal, $y_k$. The interpolation is best understood in Bayesian terms: we
compute the most probable value of $y_k$ by choosing the value that maximises its posterior probability, given
the data and a model for the signal. We choose to estimate the signal and noise separately, so  for the signal
estimate we will marginalise over the (unknown but assumed constant) noise variance.

We start by noticing that we can express the signal Fourier transform $H_k(f)$ (using either the sinc or Fresnel
approximation for this expression) as the product of our unknown signal amplitude, $y_k$ and a known signal
shape function, $\mu_k$, defined as
\be
\mu_k(f) \defeq \frac{H_k(f)}{y_k}.
\ee
If the Fourier transform of the data is $S_k(f)$ then, writing $S_{ki} \equiv S_k(f_i)$ and $\mu_{ki} \equiv \mu_k(f_i)$
the likelihood of the set of data $\{S_{k}(f)\}$ given $B_k$, spectral noise $\sigma_F$ and signal shape
$\mu_{k}(f)$ is
\be
\p\left(\{S_{k}(f)\}|B_k,\sigma_F,\{\mu_{k}(f)\}\right) = \frac{1}{(2\pi \sigma_F^2)^N}\exp\left(-\frac{1}{2\sigma_F^2}\sum\limits_{i=1}^N |S_{ki}-B_k\mu_{ki}|^2\right).
\label{eqn:SfgivenBksigmaFmuf}
\ee
where the sum in $i$ is over a window of $N$ frequency bins around the signal frequency for which $|\mu|$ is
significantly greater than zero. Here we have assumed that the noise is uncorrelated between frequency bins
and has a constant standard deviation $\sigma_F$. We can consider $\sigma_F$ as a nuisance parameter, and
marginalise over it. Choosing a Jeffreys prior for $\sigma_F$ of ${\p(\sigma_F)\propto 1/\sigma_F}$,
${\sigma_F>0}$ and a uniform prior on $B_k$, for ${B_k=-\infty}$ to $\infty$ we obtain, after marginalisation,
a log posterior for $B_k$ of
\be
\log\left[\p\left(B_k |\{S_{ki}\},\{\mu_{ki}\}\right)\right] \propto -N\log\left( \sum_i |S_{ki}-B_k\mu_{ki}|^2 \right).
\label{eqn:SfgivenBkmuffinal}
\ee
The maximum of this log posterior occurs when  ${\sum_i |S_{ki}-B_k\mu_{ki}|^2}$ is minimised. If we differentiate with respect to $B^*_k$
\footnote{$B^*_k$ has the simultaneous properties of
\begin{enumerate}
\item $B_k \rightarrow B^*_k$ is conjugate conformal, leading to $\frac{\d B^*_k}{\d B_k}=0$ and $\frac{\d B_k}{\d B^*_k}=0$
\item $B^*_k$ and $B_k$ are mutually defined; the most likely value of $B^*_k$ defines the most
likely value of $B_k$.
\end{enumerate}}
and set this to be zero we get
\begin{eqnarray}
\fl\frac{\d}{\d B^*_k}\left(\sum\limits_i |S_{ki}-B_k\mu_{ki}|^2\right) = \frac{\d}{\d B^*_k}\sum\limits_i \left(S_{ki}S^*_{ki}-B_k\mu_{ki}S^*_{ki}-S_{ki}B^*_k\mu^*_{ki}+B_kB^*_k\mu^*_{ki}\mu_{ki}\right) \nonumber\\
=\sum\limits_i \left(-S_{ki}\mu^*_{ki}+B_k\mu^*_{ki}\mu_{ki}\right)=0.
\label{eqn:dChidBk}
\end{eqnarray}
The most probable value of $B_k$ is therefore
\be
B_k=\frac{\sum\limits_i [S_{ki}\mu^*_{ki}]}{\sum\limits_i [\mu^*_{ki}\mu_{ki}]},
\label{eqn:LSEstimator}
\ee
a result that is familiar from least-squares analysis.

To estimate the variance of the noise, $\sigma_k^2$, we would ideally follow a similar route, marginalising
over $B_k$ in \eref{eqn:SfgivenBksigmaFmuf} and maximising the posterior for $\sigma_k$
\begin{equation}
\p(\sigma_k|\{S_{ki}\},\{\mu_{ki}\})_{\rm max}=\left[\int\limits_{-\infty}^{\infty}\p(\sigma_k|\{S_{ki}\},\{\mu_{ki}\},B_k)\p(B_k) \d B_k\right]_{\rm max}.
\label{eqn:marginBk}
\end{equation}
However, this integral is not analytic. We therefore choose to use our calculated value of $B_k$ from
\eref{eqn:LSEstimator} to obtain the best estimate of $\sigma_k$, equivalent to using the Dirac delta function
as the prior on $B_k$ in \eref{eqn:marginBk}
\be
\p(B_k) = \delta_D\left(B_k-\frac{\sum\limits_i [S_{ki}\mu^*_{ki}]}{\sum\limits_i [\mu^*_{ki}\mu_{ki}]}\right).
\ee
The application of this is straightforward: we use the most probable $B_k$ calculated above to return best
estimate of $H_k(f_i)$, ${H_{\rm best}(f_i)=B_k \mu(f_i)}$. The difference between $H_{\rm  best}(f_i)$ and
$S(f_i)$ is our best estimate of the noise $N_{\rm best}(f_i)$. We take these noise residuals around the signal
frequency and then calculate their variance to give us $\sigma_F^2$.

The spectral noise variance, $\sigma_F^2$, is related to the time domain noise through Parseval's theorem
($\sigma_F^2 = \sigma_T^2 \frac{2}{\Delta t^2 r}$). Using this and \eref{eqn:noiseTDtokconvert} we get
\be
\sigma_k^2 = \sigma_F^2 \frac{\Delta t}{2}.
\label{eqn:noiseFDtokconvert}
\ee
We now have our calculated $B_k$ and an estimate of $\sigma_k$. The parameter estimation stage of the pipeline
used by Dupuis and Woan \cite{RJDandWoan} treated the noise variance as a nuisance parameter to be marginalised
over 30-minute segments to give a Student's~$t$-likelihood for the signal. However here we use the direct
estimates of $\sigma_k$ described above, giving a Gaussian likelihood for use in parameter estimation.

\subsection{Outlier removal}
\label{SubSec:OutlierRemoval}
The noise in gravitational-wave data contains many line features that may affect our estimates of $B_k$ if they
are close to the source frequency. We minimise this contamination by performing three outlier removal steps.
The first outlier removal routine uses the standard deviation of $S(f)$ as an initial estimate of the noise. We 
then multiply this standard deviation by a number (typically around ten) decided by the user and remove any $S(f)$ data 
points with an absolute value above this threshold. This threshold is
set to be large, to remove only the strongest lines, and the five bins either side of the signal frequency are excluded
from this first step.

\begin{figure}[t]
\centering
\includegraphics[width=0.7\linewidth]{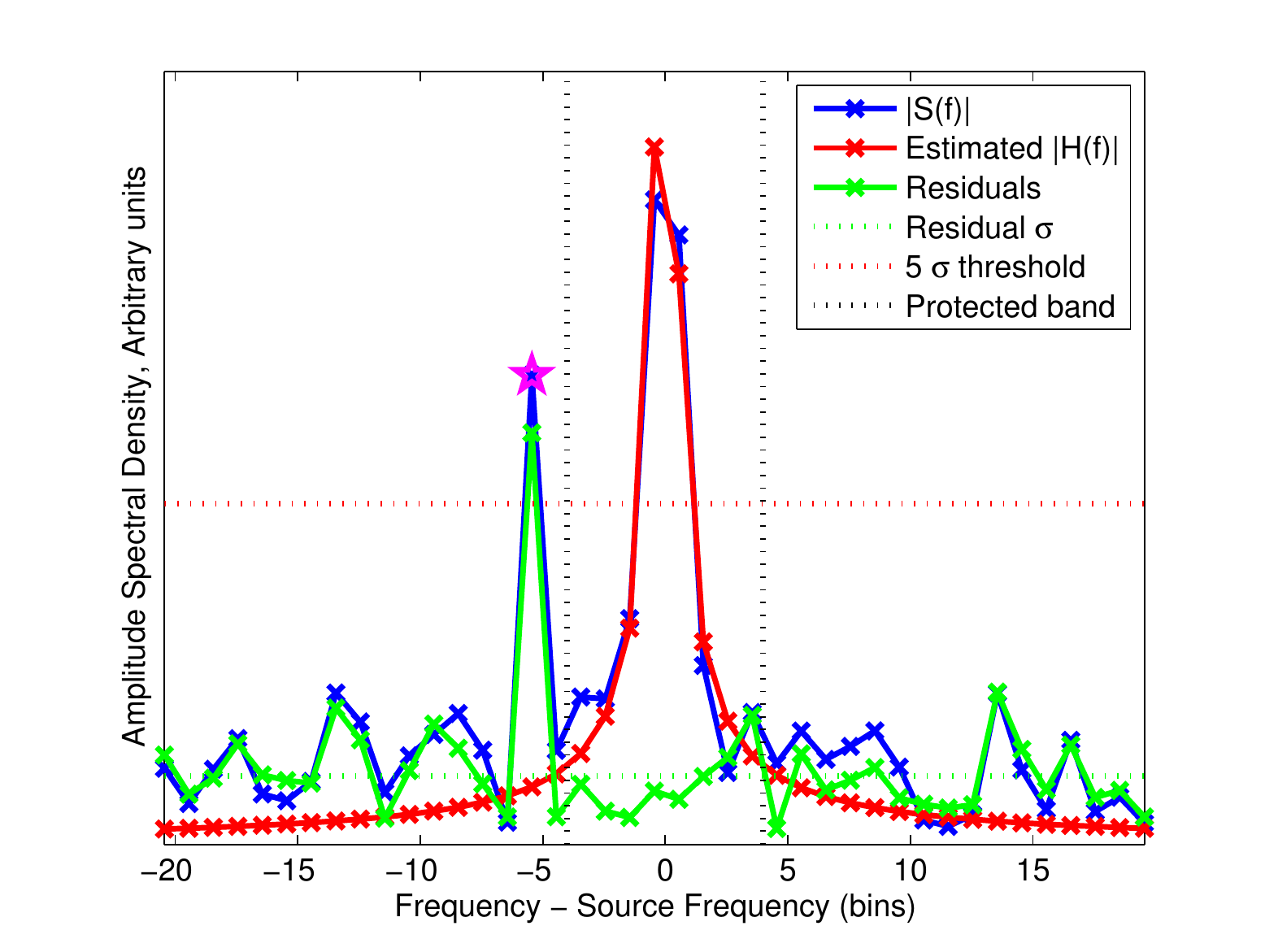}
\caption{An illustration of the type of outlier removed by
the second outlier removal step. Shown are the best fit of the noise, the
standard deviation of the residuals, the threshold for removal, and the
protected band around the source frequency. The removed data point is
indicated by the magenta star.}
\label{fig:OR2Example}
\end{figure}

The second outlier removal step takes place after initial estimates of $B_k$ and $\sigma_k$ have been made and is shown in
\fref{fig:OR2Example}. By this stage we have an estimate of the noise in the frequency domain, $\sigma_F$, so
we identify $S(f)$ data points with residual values above a threshold factor of this standard deviation. We use a
factor of five in the illustrations given here. This threshold is lower than that employed in the first step,
and now all but the closest $\pm 4$ data points to the signal are involved. If any data points are removed by
this process $B_k$ and $\sigma_F$ are recalculated and this outlier removal step repeated to convergence.

\begin{figure}[t]
\centering
\includegraphics[width=0.7\linewidth]{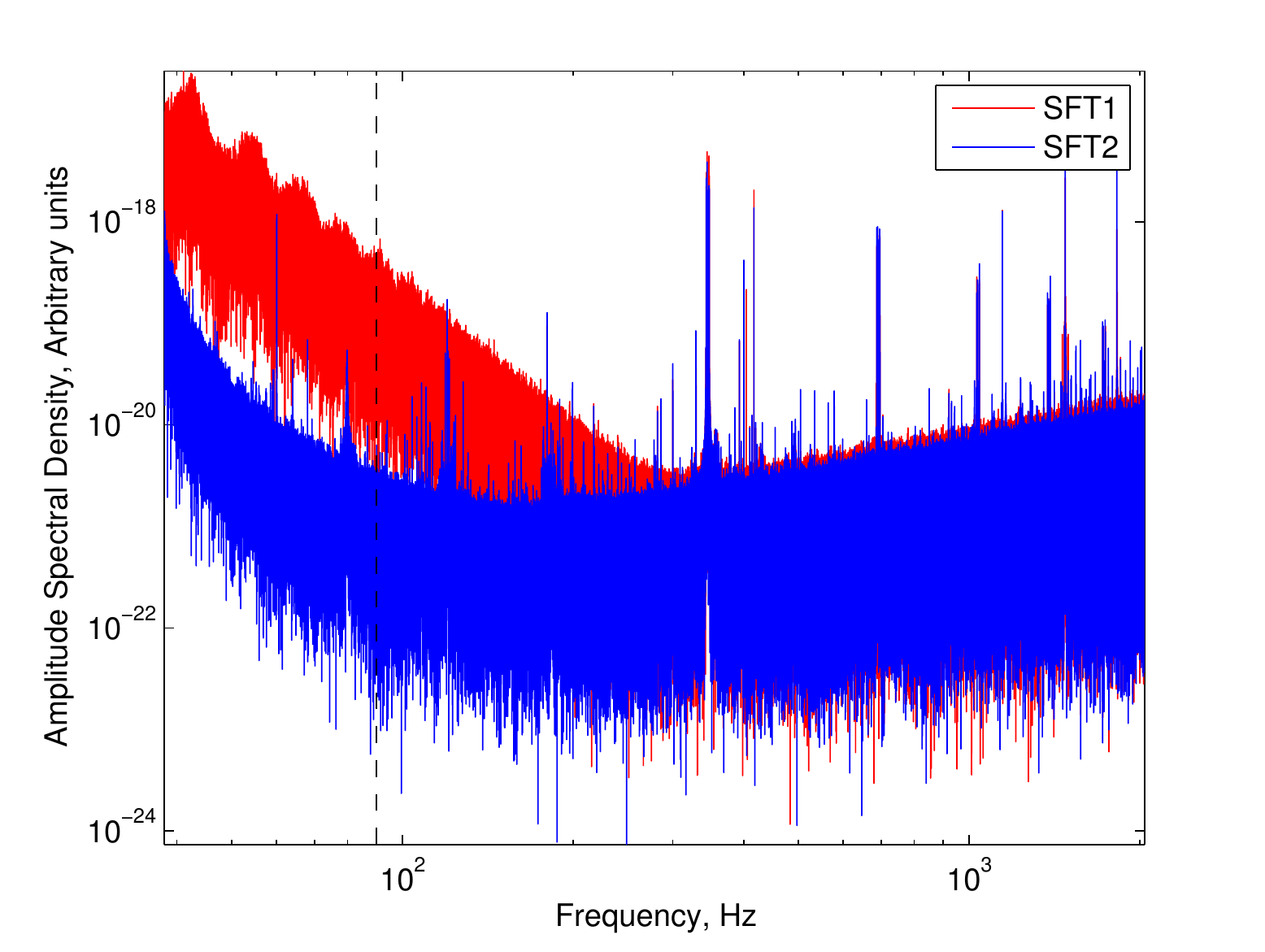}
\caption{An illustration of the type of outlier removed by the
third outlier removal step, showing the power spectra of two example SFTs.
SFT1 has an unusually high low-frequency noise contribution, bleeding power into
frequency channels up to around 300\,Hz. For sources with signals in this lower range
(such as a 90-Hz signal shown by the black line) the noise estimation for SFT1 would be large compared to that in SFT2
(a normal SFT). Sources at frequencies above $\sim 300$\,Hz would be unaffected by this outlier removal.
}
\label{fig:OR3Example}
\end{figure}

The third outlier removal step takes place after all $B_k$ and $\sigma_k$ have been calculated, and uses the noise estimates over the entire data set. We calculate the mean value
of $\{\sigma_k\}$, $\langle \sigma \rangle$, and remove any data for which $|\mathfrak{Re}[B_k]|$,
$|\mathfrak{Im}[B_k]|$ or $\sigma_k$ is above a threshold factor of this value. This step removes $B_k$ data points
which are unusually noisy, but for which the noise is broadband and was not detected by the first two outlier
removal steps, as shown in \fref{fig:OR3Example}.

We consider it rather unlikely that our methods would accidentally veto astrophysical signals that are slightly offset from our expected frequency, 
as a real signal would have to be exceptionally strong to show up as significantly as the lines we are vetoing in a single SFT.

\begin{figure} [h!]
\centering
\includegraphics[width=0.7\linewidth]{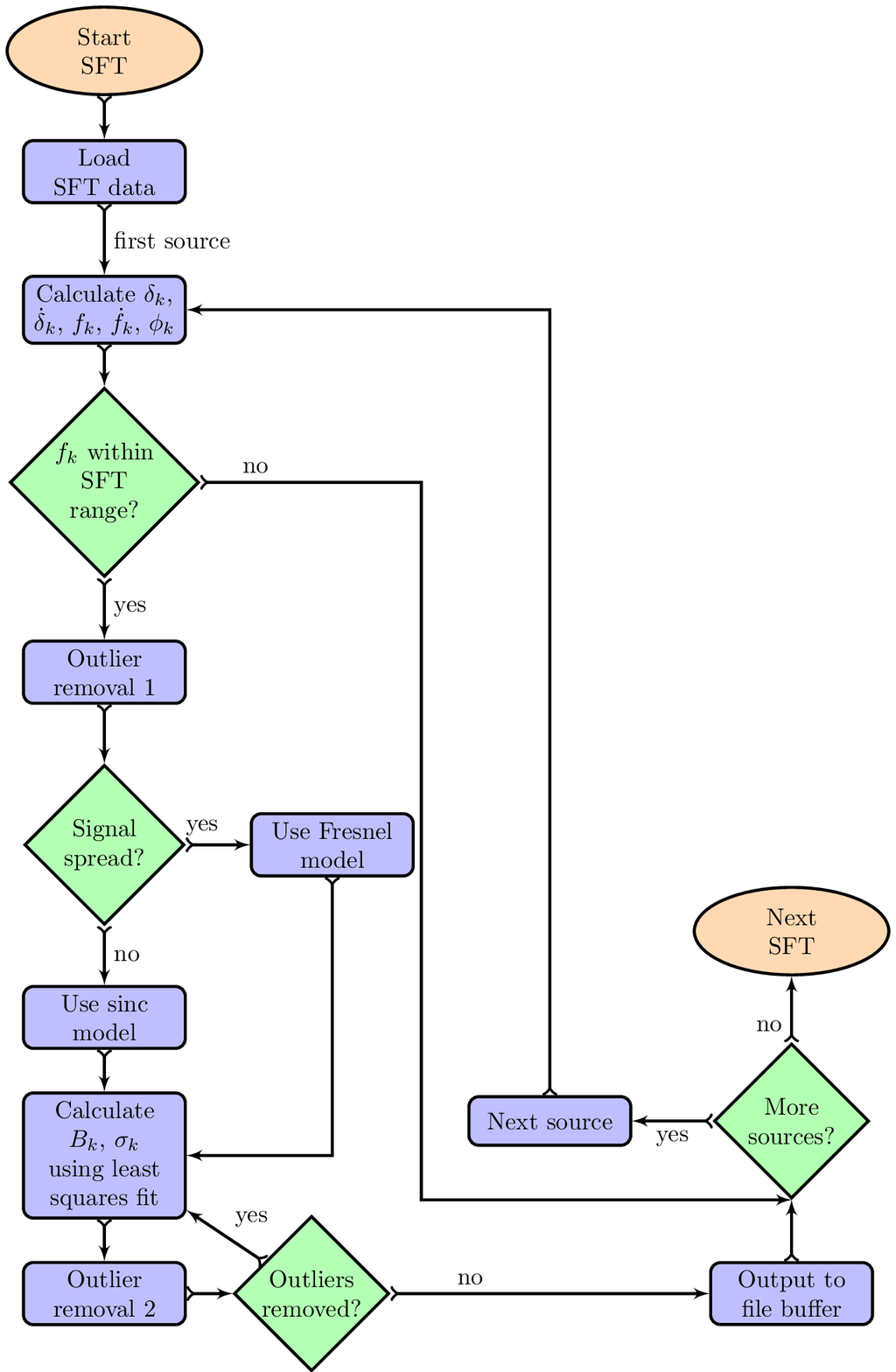}
\caption[The Spectral Interpolation algorithm.]{Flowchart showing the Spectral Interpolation algorithm
during each SFT. This flowchart runs for each SFT, running on a loop within each science segment, which
itself is looped over. The third outlier removal is not shown, as it does not take place within this loop.}
\label{fig:SplAlgFlowchart}
\end{figure}

Figure~\ref{fig:SplAlgFlowchart} shows the full SplInter algorithm for a single SFT. The detector data includes
data quality flags and we restrict our analysis to segments of data in `science mode'. The input files are
therefore a science segment list and pointers to the corresponding Fourier data and a set of files defining the
targets. The SplInter algorithm loops through segments, and in each segment processes each SFT according to
\fref{fig:SplAlgFlowchart}.

\section{Testing the SplInter algorithm}
\label{Sec:SplTest}
We tested the SplInter algorithm against the standard heterodyne method currently employed for both accuracy
and performance. The first accuracy test is described in \sref{SubsubSec:Puresignal} and checked that the
$B_{k/K}$ outputs from the two routines are consistent in the noiseless case. In
\sref{SubsubSec:NoiseTest} we check that the noise estimation $\sigma_k$ is also accurate, and that this
estimate is consistent with that estimated from the heterodyned $B_K$. In \sref{SubsubSec:HWInj} we
perform  a black-box replacement test, comparing the performance of our routines end-to-end for the analysis of
hardware signal injections in LIGO S6 data \cite{0264-9381-32-11-115012}. Finally we  test the algorithm performance in
\sref{SubSec:SpeedTest}, particularly the speed increase of SplInter compared to the heterodyne routine.

\subsection{Accuracy and testing}
\label{SubSec:Accuracy}
We now compare the SplInter output, $B_k$, with the standard heterodyne output, $ B_K$ (which we assume
to be exact for this comparison), using the \textit{mismatch}, $m$, between the two, defined as
\be
m = \abs{1 - \frac{\sum_k B_{k{\rm, Spl}} \cdot B_{k{\rm, het}} }{\sum_k B_{k{\rm, het}} \cdot B_{k{\rm, het}}}}.
\label{eqn:mismatch}
\ee
The mismatch is a useful indicator of how well our approximation matches the exact solution, and gives an approximate 
figure for the drop in SNR caused by these approximations.
We define $B_{k{\rm, het}}$ as the average $B_K$ value over the duration of the corresponding SFT, which is
equivalent to performing the heterodyne with a $\Delta t$ value of 30 minutes.

\subsubsection{Recovery of noiseless signals from isolated pulsars}
\label{SubsubSec:Puresignal}
\begin{figure}
\centering{
\includegraphics[width=0.7\linewidth]{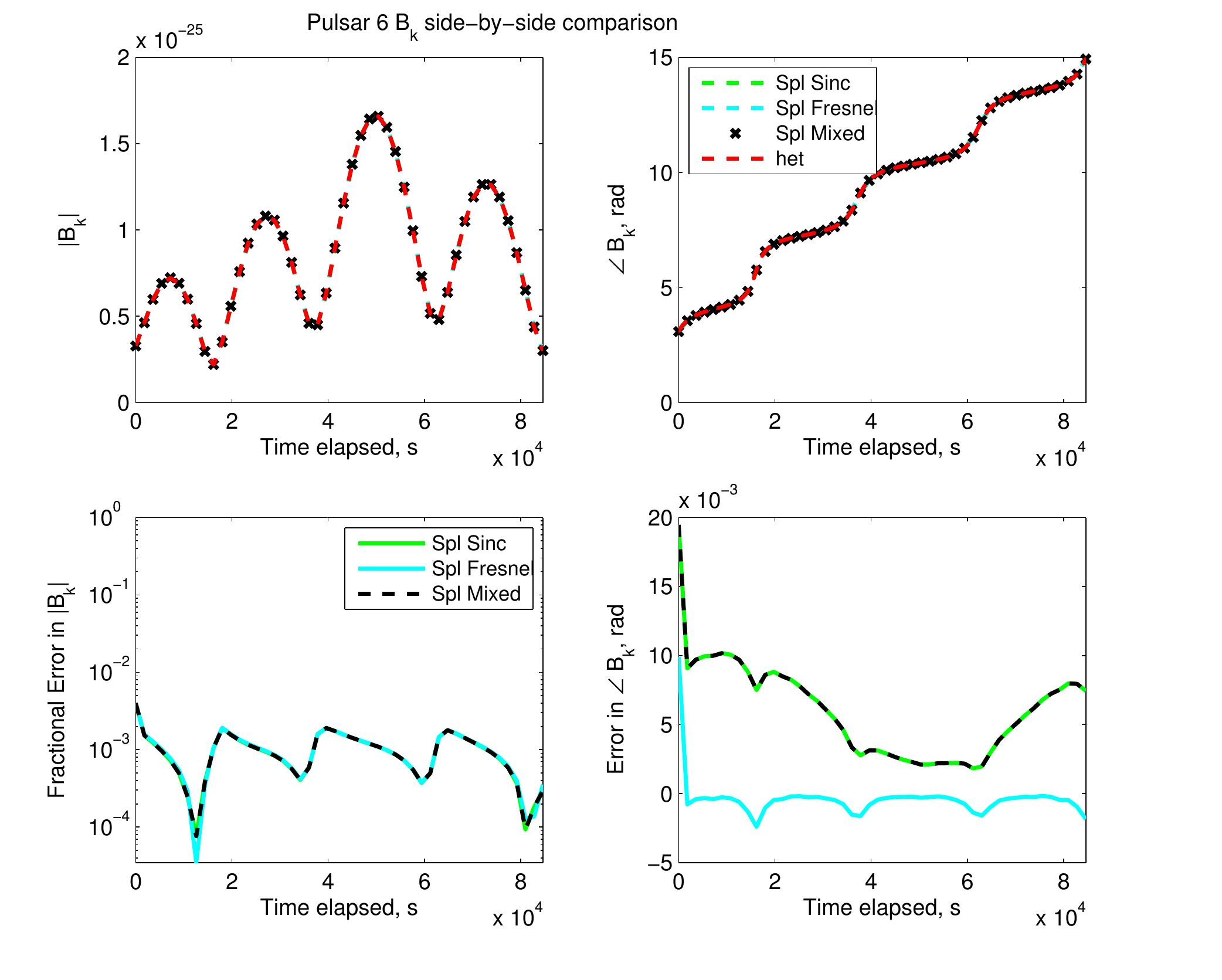}
}
\caption{SplInter $B_k$ (green/cyan/black) and fine heterodyne $B_K$ (red) values over one day, with amplitude
on the left and phase on the right, for a noiseless signal corresponding to hardware injection PULSAR6.  Below
are the fractional difference between the two.  SplInter values are shown using the sinc approximation
(green), the Fresnel approximation (cyan) and the mixed interpolator (black), which uses the sinc
approximation when $|\dot{f}_k|\Delta t^2<0.1$ and the Fresnel approximation otherwise. $m = 0.0029 $ for the
sinc approximation, $ m = 0.0029 $ for the mixed interpolation and $m = 0.0028 $ for the Fresnel
approximation. The two methods are equally precise in this case, as the frequency does not significantly
change during the SFT length.} \label{fig:BkComp3}
\end{figure}

\begin{figure}[!h]
\centering
\includegraphics[width=\linewidth]{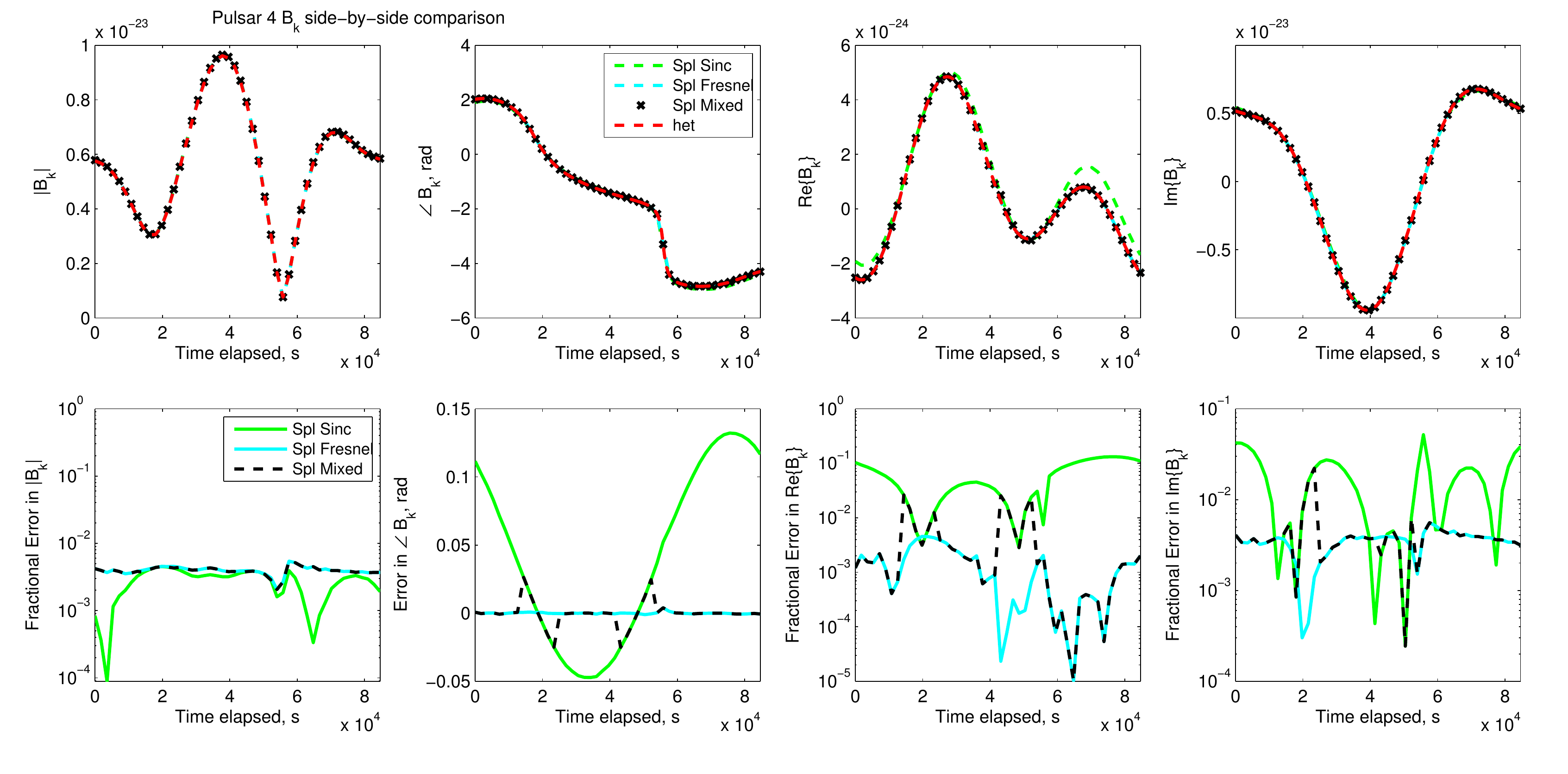}
\caption{SplInter $B_k$ (green/cyan/black) and fine heterodyne $B_K$ (red) values over one day, showing amplitude, phase and real/imaginary parts for a noiseless signal corresponding to hardware injection PULSAR4 (colours as
\fref{fig:BkComp3}. $|\dot{f}_k|$ is high for this pulsar, and the Fresnel approximation is used by the mixed interpolator to maintain accuracy. There is a significant discrepency in the real part of $B_k$ when using just the sinc approximation.
($m = 0.0101$ for sinc, $0.0040$ for mixed and $0.0039 $ for the Fresnel approximation.}
\label{fig:BkComp4}
\end{figure}

In the case of a noiseless signal, the heterodyne and spectral interpolation routines should, ignoring
approximations, agree exactly, as $B_k = y_k$. Figures~\ref{fig:BkComp3} and~\ref{fig:BkComp4} show the result
of applying the SplInter and heterodyne routines to noise-free data. The frames and SFTs were made using
\texttt{lalapps\_create\_pulsar\_signal\_frame} and \texttt{lalapps\_MakeSFTs} respectively\footnote{These
routines are within the LALsuite software repository
\url{https://www.lsc-group.phys.uwm.edu/daswg/projects/lalsuite.html}}. We see that the SplInter and
heterodyne $B$-estimates agree well, and always to better than 1\%. We apply a hybrid interpolation scheme here, using the sinc approximation when $\dot{f}_k$ is small, and the Fresnel approximation otherwise, with a changeover point at $|\dot{f}_k|\Delta t^{2} = 0.1$ \cite{GSDThesis}.

\Table{\label{tab:PSRInjParams} Parameters of hardware injection pulsars 4 and 6.}
\begin{tabular}{|c||cc|cc|}
\hline
~& $f^{(0)}$\,Hz & $f^{(1)}$\,Hz\,s$^{-1}$ & RA & Dec\\ \hline
PULSAR4 & 1403.16 & $-2.54\times10^{-8}$ & $18^{\rm h}\,39^{\rm m}\,57.04^{\rm s}$ & $-12^\circ\, 27'\, 59.85''$ \\
PULSAR6 & 148.72 & $-6.73\times10^{-9}$ & $23^{\rm h}\,55^{\rm m}\, 0.23^{\rm s}$ & $-65^\circ\, 25'\, 21.45''$ \\ \hline
\end{tabular}
\endTable


Figure~\ref{fig:BkComp4} shows the importance of using the Fresnel rather than the sinc approximation for
PULSAR4. This source has both a high frequency and a relatively low declination, leading to a large second
order change in phase over the duration of the SFT.

\subsubsection{Recovery of noiseless signals from binary pulsars}
\label{SubsubSec:PuresignalBinary}
\begin{figure}[!h]
\centering
\includegraphics[height=0.4\textheight]{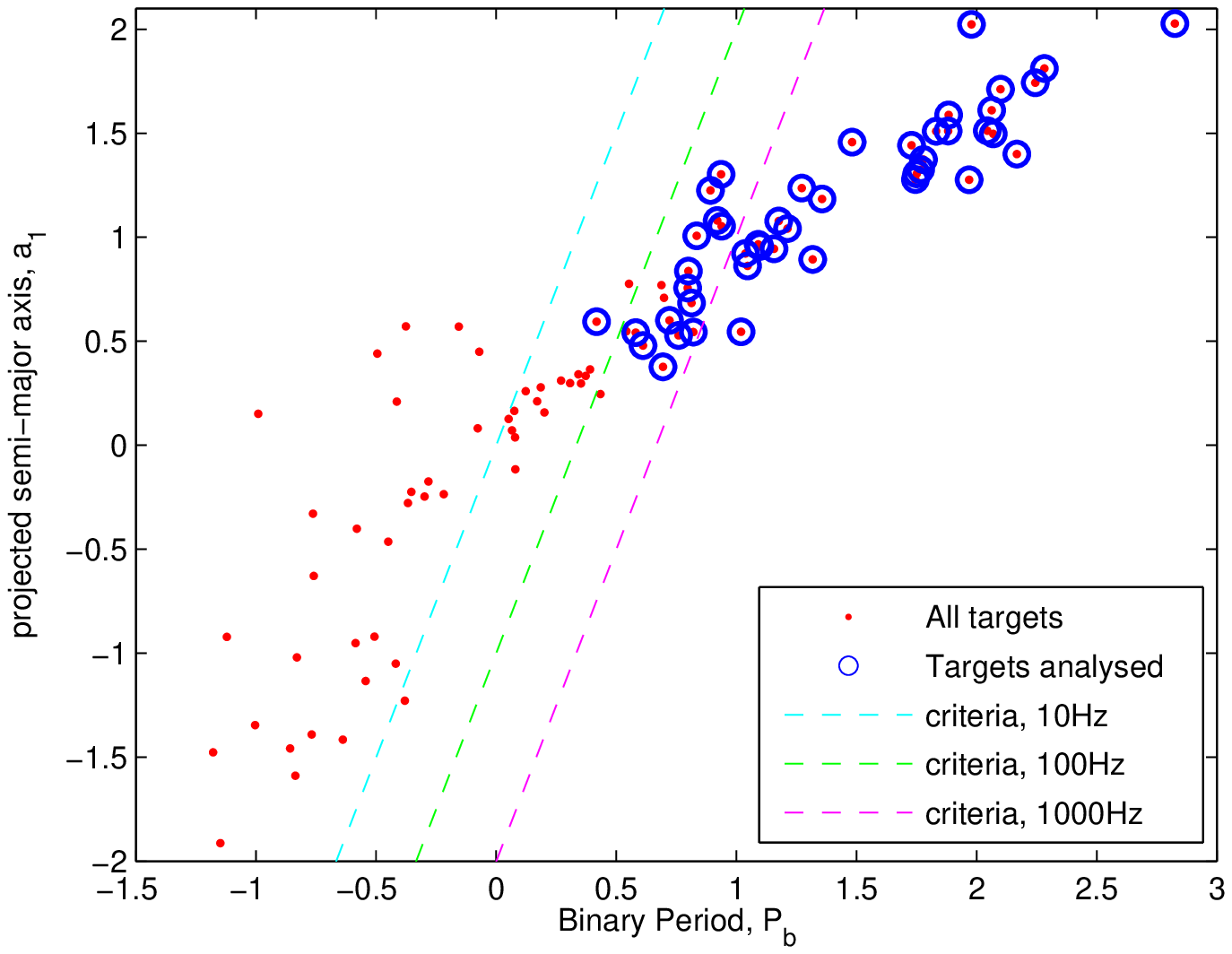}
\caption{Binary period vs projected
semi-major axis for targeted binary pulsars, indicating which binary pulsars have a small enough mismatch to be analysed using
SplInter and which cannot. We include an indication of the empirical criteria we set for analysis of a target
in a binary system, given in \eref{eqn:binCrit}, for pulsars with source frequency of 10, 100 and 1000\,Hz.}
\label{fig:PbvsA1}
\end{figure}

The signal delay for a binary pulsar contains an extra term,  $\Delta_{\rm Binary}$ , in \eref{eqn:deltat}
from the Roemer, Shapiro and Einstein delays in the binary system itself, and this delay can introduce rapid
variations in apparent frequency. Figure \ref{fig:PbvsA1} shows which of the 97 known binary pulsars have a mismatch below 
0.1 (circled) when comparing the $B_k$ values analysed with SplInter and heterodyne respectively.
We compute the mismatch over one day if the binary
period $P_{\rm b}$ is $<$1\,d, over the binary period if $1<P_{\rm b}<5$\,d and over 5\,d if $P_{\rm b}>5$\,d.

50 of these sources are in systems that show mismatches above 0.1 for SplInter, and we are therefore unable to
use this method for these and maintain accuracy and sensitivity. The mismatch comes from significant
high-order frequency derivatives in these pulsars over the 30\,min period of the SFT. The second-order
frequency derivative $\ddot{f}_k$ is proportional to $f^{(0)}a_1/P_{\rm b}^3$, where $a_1$ is the projected
semi-major axis of the binary system. By considering which binary pulsars we are unable to analyse using SplInter 
(those not circled in \fref{fig:PbvsA1}), we can set an empirical upper-limit on $\ddot{f}_k$. When using 30\,min SFTs, 
this limit is 
\be
\frac{f^{(0)}}{1\,\rm{Hz}}  \frac{a_1}{1\,{\rm light second}} \left(\frac{P_{\rm b}}{1\,{\rm day}}\right)^{-3}\lesssim10,
\label{eqn:binCrit}
\ee
and this limit is delineated in \fref{fig:PbvsA1} for pulsars with gravitational wave frequencies of 10, 100
and 1000\,Hz.

\subsubsection{Noise estimation tests}
\label{SubsubSec:NoiseTest}
We tested noise estimation using SFTs and frames with known levels of white noise but no signal. After running
the SplInter and heterodyne algorithms, we checked that the $B_{k/K}$ noise estimates were consistent with the
injected value and with each other. We compared noise estimates from the SplInter routine to noise estimates
from the heterodyne routine for noise data with a time-domain variance of $\sigma_T^2=1$. The estimate of the
noise from the heterodyne, $\sigma_H$, was obtained using an average of the variance of the real and imaginary
heterodyne parts over the duration of an SFT, converted into the equivalent noise value for the 30-min cadence
of $B_k$.

\begin{figure}[h!]
\centering
\includegraphics[width=0.7\linewidth]{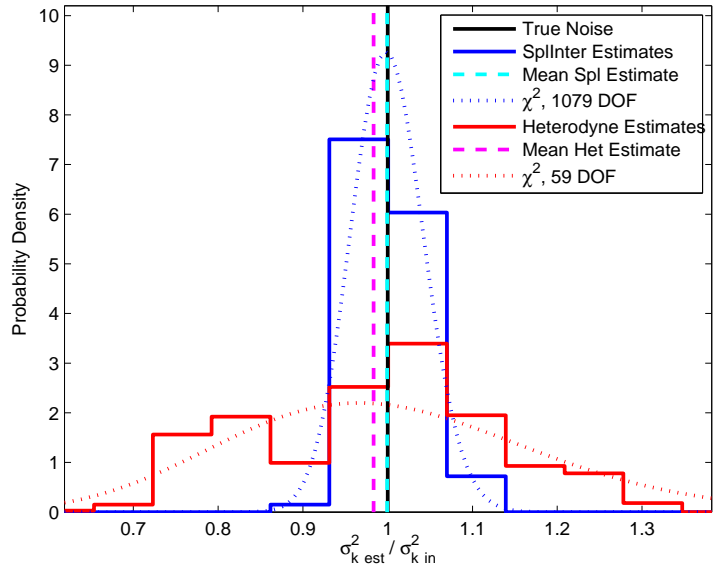}
\caption{A histogram of standard deviation estimates of white noise from SplInter (top) and heterodyne (bottom), with the
mean estimated values shown as dashed lines. Also shown is the true value of the noise (black dashed line) and
the expected distributions of these values (dotted lines).}
\label{fig:noiseEstimates}
\end{figure}

We see in \fref{fig:noiseEstimates} that the heterodyne and SplInter noise estimates are consistent, and that
both agree with the injected value of the noise and the expected distribution of these estimates. The expected
distribution is a $\chi^2$ distribution with $n-1$ degrees of freedom, where $n$ is the number of data points
used to calculate the noise estimate:
\be
 \frac{\sigma_k^2}{\sigma_{\rm true}^2} \sim \frac{\chi^2(n-1)}{n-1},
\ee
which for large $n$ approximates a normal distribution with unit mean and variance $2/(n-1)$.
The heterodyne noise estimate used 30 $B_K$ data points from each of the real and imaginary parts of the data,
leading to an expected distribution of a $\chi^2$ with 59 degrees of freedom, shown in the figure by the red
dotted line. Here we used the spectral interpolation algorithm with a bandwidth of 0.3\,Hz around the signal
frequency, leading to a $\chi^2$ distribution on $\sigma_k^2$ with 1079 degrees of freedom (shown on the
figure as a blue dotted line). One might get a marginally better noise estimate using a wider bandwidth,
however the frequency dependence of the noise limits this. Additionally,  there are diminishing returns in
computational efficiency, and repeated use of the algorithm has found that a bandwidth of 0.3Hz is a good 
compromise between these considerations.
\subsubsection{Full testing with hardware injections}
\label{SubsubSec:HWInj}
The LSC and Virgo collaborations inject artificial signals into the detector hardware control loops to test
analysis pipelines.  Here we show the results of running both the heterodyne and SplInter pipelines on two
hardware-injected pulsars ( `PULSAR4' and `PULSAR6', the parameters of which are given in
\tref{tab:PSRInjParams}) in just under four months of LIGO S6 data from the Hanford detector (LHO).  During
this interval the LHO duty cycle was 47\%, giving $\sim 5\times 10^6$\,s of science data. With 4 months of
data the injections can be recovered with a high signal-to-noise ratio, but with the posteriors retaining
sufficient width to usefully assess our noise estimates. \Tref{tab:ShortHWInjSNRs} lists the returned
signal-to-noise ratios for the two pulsars using both the SplInter and heterodyne pipelines, running the
latter with both a Gaussian and Student's~$t$ likelihood.

\Table{\label{tab:ShortHWInjSNRs}SNRs of hardware injections 4 and 6 recovered by the two pipelines from
four months of LIGO Hanford S6 strain data, calculated using the nested sampling algorithm
\texttt{lalapps\_pulsar\_parameter\_estimation\_nested} \cite{2012JPhCS.363a2041P}}
\begin{tabular}{|c||c|c|c|}
\hline
  $B_{k/K}$ algorithm & SplInter & \multicolumn{2}{|c|}{heterodyne} \\ \hline
  Likelihood distribution & \multicolumn{2}{|c|}{Gaussian} & Student's~$t$ \\ \hline
  PULSAR4 & 234.67 & 251.99 & 235.23 \\
  PULSAR6 & 16.02 & 17.22 & 17.18 \\ \hline
\end{tabular}
\endTable

We see that the SplInter results are consistent with those from the heterodyne pipeline, with SNR values
around 7\% below those from heterodyne. This is to be expected in real data containing segments and lines as
the filtering is different in the two pipelines. Most of this drop in SNR (around 5\%) is due to SplInter's
need for contiguous 30-min stretches of science data, rather than the 60\,s stretches used by the heterodyne
pipeline.

\begin{figure}[h!]
\centering
\includegraphics[width=0.7\linewidth]{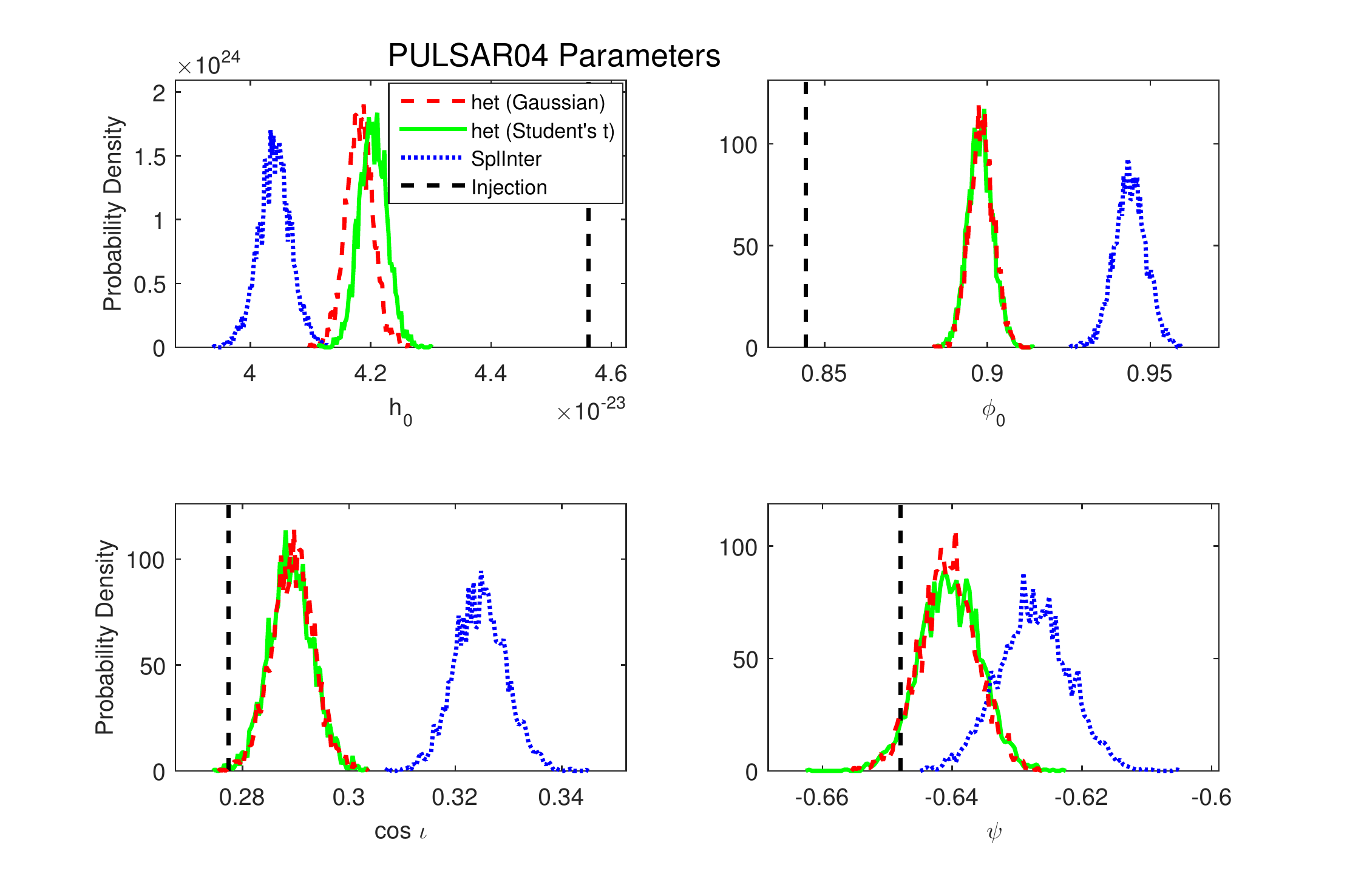}
\caption{Posterior distributions on $h_0$, $\phi_0$, $\cos\iota$ and $\psi$ for hardware injection PULSAR4
using  four months of LHO S6 data. The blue dotted line shows posteriors made using SplInter for the calculation of
$B_{k/K}$, and the red dashed and green solid lines show the heterodyne algorithm with Gaussian and
Student's~$t$-likelihood distribution respectively. The vertical black line shows the injection value, which is 
slightly offset from the recovered signal due to calibration uncertainties.}
\label{fig:ShortHWInjPosts3}
\end{figure}

\begin{figure}[h!]
\centering
\includegraphics[width=0.7\linewidth]{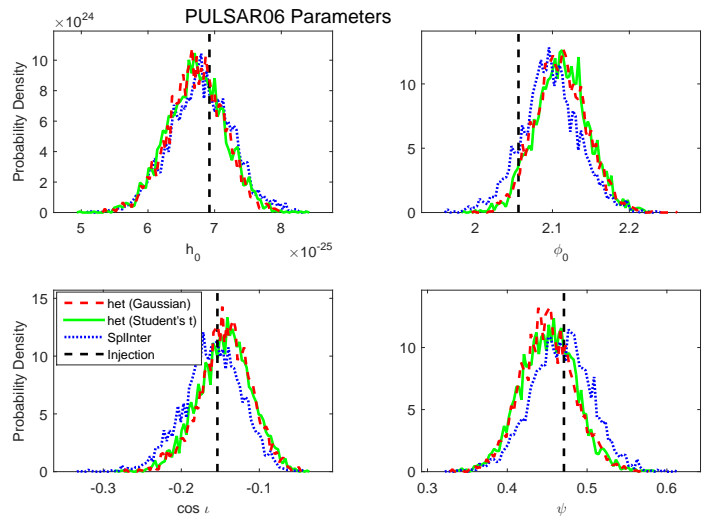}
\caption{Posterior distributions on $h_0$, $\phi_0$, $\cos\iota$ and $\psi$ for hardware injection PULSAR6
using four months of LHO S6 data. Colour scheme as in \fref{fig:ShortHWInjPosts3}.}
\label{fig:ShortHWInjPosts6}
\end{figure}

Figures~\ref{fig:ShortHWInjPosts3} and~\ref{fig:ShortHWInjPosts6} show the posterior
distributions of the four source parameters determined in targeted searches, $h_0$, $\phi_0$,
$\cos\iota$ and $\psi$.
Again, in this example the posteriors generated by the two pipelines show sufficiently good agreement to allow
the SplInter pipeline to replace the heterodyne pipeline without a systematically significant impact on overall
performance. In \fref{fig:ShortHWInjPosts3} the discrepancies between the injected values and the recovered
posteriors are within the range expected due to calibration uncertainties, and the discrepancies between the heterodyne 
and SplInter data are small enough to demonstrate that SplInter is a viable replacement for most targets, but that 
the heterodyne method should be retained for more accurate analysis.

\subsection{Speed testing}
\label{SubSec:SpeedTest}
The purpose of SplInter is to decrease the computational cost of targeted searches without significantly
affecting sensitivity. We now consider the speed of SplInter in comparison with the heterodyne algorithm.
The fundamental speed increase comes from the fact that we do not require the entire data bandwidth for our
estimate of $B_k$. SplInter only requires a small frequency band of less than 1\,Hz, whereas the heterodyne
algorithm initially starts with a dataset containing the equivalent of 16384\,Hz. In addition to this, we can
analyse the sources in parallel for each SFT, reducing overall file input/output time when alaysing multiple sources.

It is simplest to compare the total algorithm time taken per source, as the heterodyne algorithm simply takes
the sources one at a time. However the SplInter execution time is not linear in the number of sources, so we
also compare the total time per source, for one, ten, one hundred and one thousand sources at a time. These
tests use the mixed interpolation scheme, so that we gain accurate timing results, including the occasional
use of the Fresnel approximation.

\begin{figure}[h!]
\centering
\includegraphics[width=0.7\linewidth]{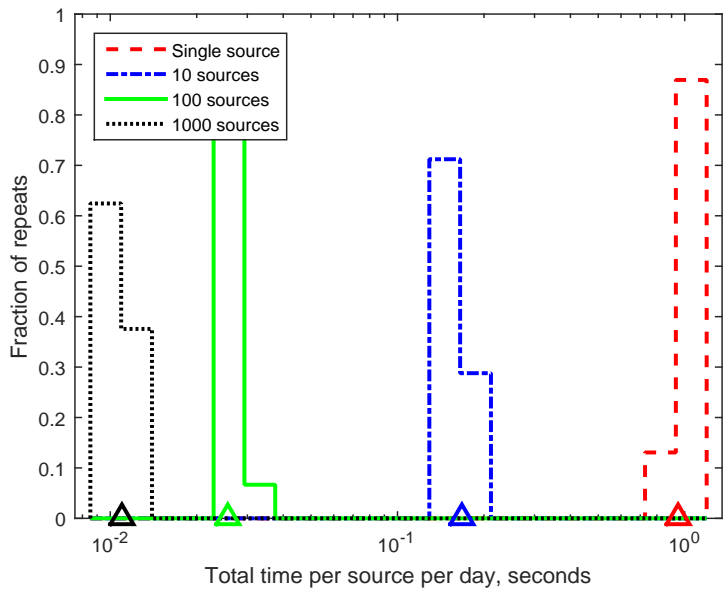}
\caption{Histograms of the average time taken to analyse sources for a day of continuous data for the SplInter
algorithm. The time taken by the heterodyne algorithm is 5.0$\times 10^{2}$ seconds per day per source, which 
would be over two orders of magnitude past the right hand limit of this figure. The different coloured histograms show the different
number of sources used in each analysis. The horizontal axis markers denote the mean values for these times
for the different numbers of sources. }
\label{fig:TimeTotal}
\end{figure}

\Tref{tab:totalTimingStats} and \fref{fig:TimeTotal} show the amount of time taken to analyse the sources for
a day of continuous data using both the heterodyne and SplInter routines. \Tref{tab:totalTimingStats} also
shows the computational improvement in CPU hours per source per hour of data. We see that the SplInter routine
can improve the computational efficiency of the $B_{k/K}$ calculation by up to two orders of magnitude for
single source input, and four orders of magnitude if we use multiple source input. These analyses were
performed on the \texttt{atlas} computing cluster at the Albert Einstein Institute, Hannover.

\Table{\label{tab:totalTimingStats}Median time taken to analyse sources for a day of data using heterodyne and
Spectral Interpolation in seconds and CPU core hours per number of sources per hour of data for each
interferometer.}

\begin{tabular}{|c|c|c|c|c|c|}
\hline
  &  Heterodyne     & \multicolumn{4}{|c|}{SplInter} \\ \hline
 Sources  & 1                   & 1                   & 10                  & 100                  & 1000 \\ \hline
 Time/day (s) & 5.0$\times 10^{2}$  & 1.0                 & 1.7                 & 2.5                  & 10.9 \\
 CPUh/N/h & 5.8$\times 10^{-3}$ & 1.1$\times 10^{-5}$ & 1.9$\times 10^{-6}$ & 2.9$\times 10^{-7}$ & 1.2$\times 10^{-7}$ \\ \hline
\end{tabular}
\endTable
The improvement in computational efficiency is not just limited to the $B_k$ calculation stage. The lower
cadence of the SplInter $B_k$ data results in fewer data points containing a similar amount of information and
hence faster calculations at the parameter estimation stage. This does however come with a cost: SplInter
requires contiguous 30-min periods and lacks the flexibility of the heterodyne code in that respect.

\section{Discussion}
We have introduced \textit{SplInter}, a new spectral interpolation method of calculating the down-sampled
complex amplitude of a continuous wave signal with relative motion and source rotation effects removed. We
have shown that this algorithm improves the computational efficiency of this part of the Bayesian targeted and
$\mathcal{G}$-statistic pipelines by up to four orders of magnitude, and have explained how using longer time
steps for $B_{k/K}$ with an estimate of the noise has a knock-on effect of improving computational speed in
the parameter estimation stage. We have also shown that the SplInter routine performs well in comparison to
the heterodyne routine in most cases, and that there is no significant drop in the recovered SNR. The increase
in computational efficiency means that the search is a viable rapid follow-up pipeline for all-sky or directed
search candidates, and preliminary results for this secondary use are in \cite{GSDThesis}. This method has
been used in dual-harmonic searches for gravitational waves from spinning neutron stars
\cite{2015arXiv150800416P}, in which the Spectral Interpolation algorithm was found to improve the upper limit
on J1748$-$2446ac by a factor of 1.7 compared to \cite{2010ApJ...713..671A} by the use of the line removal
routine.

Tests on the $B_k$ output of the SplInter and heterodyne output have shown that the SplInter algorithm is not 
suitable for sources in binary systems with relatively short binary periods compared to the length of the SFT, this is as
the frequency of the signal will alter significantly and non-linearly over the course of the SFT. Work is
planned to provide a solution to this problem, which could include switching between the time and frequency
domain to make shorter Fourier transforms in the cases of high frequency variability during the 30\,min
duration of the SFTs. The inverse FFT and FFT required to do this would not be computationally expensive, due
to the efficiency of the FFT and inverse FFT algorithms. This method could be able to be used in a flexible
way, calculating the required timestep for the new transforms for each SFT, meaning that the number of
returned data points is optimal.

\section{Acknowledgements}
The authors would like to thank members of the Institute for Gravitational Research at the University of
Glasgow and the LIGO Scientific Collaboration continuous waves group for constructive comments and discussion.
In particular we would like to thank Damir Buskulic for useful comments in preparation of this manuscript. 
GSD is funded by a studentship from the Science and Technologies Facilities Council. MP and GW are funded by
the STFC through grant number ST/L000946/1. This document has been assigned the LIGO Document Control Center
number LIGO-P1500258.

\section*{References}
\bibliographystyle{unsrt}
\bibliography{BibSplPaper}
\end{document}